\documentclass[useAMS,usenatbib]{mn2e}
\usepackage{graphicx}

\title[Damping of neutron star shear modes by superfluid friction]
{Damping of neutron star shear modes by superfluid friction}
\author[P. B. Jones]
       {P. B. Jones \thanks{E-mail: p.jones1@physics.ox.ac.uk}\\
        University Of Oxford, Department of Physics, Denys Wilkinson Building,
		Keble Road, Oxford, OX1 3RH\\}
\date{}

\pagerange{\pageref{firstpage}--\pageref{lastpage}}
\pubyear{}

\begin{document}

\maketitle

\label{firstpage}

\begin{abstract}
The forced motion of superfluid vortices in shear oscillations of rotating
solid neutron star matter produces damping of the mode. A simple
model of the unpinning and repinning processes is described, with
numerical calculations of the consequent energy decay times.  These are
of the order of 1 s for typical anomalous X-ray pulsars with rotational
angular velocity $\Omega \approx 1$ rad s$^{-1}$ but become very
short for radio pulsars with $\Omega \approx 10 - 100$ rad s$^{-1}$. The
superfluid friction processes considered here may also be significant
for the damping of r-modes in rapidly rotating neutron stars. 
\end{abstract}


\section{Introduction}

Although there has been much recent interest in shear modes of neutron
star oscillation (Blaes et al 1989; Thompson \& Duncan 1995; Duncan 1998),
relatively little attention has been paid to the associated damping
mechanisms apart from coupling with magnetospheric Alfv\'{e}n
waves.  Blaes et al consider that stress-induced
motion of dislocations may be an important process, as is
believed to be the case for shear waves in the Earth.
But nothing is known about the distribution of linear
and planar defects in the lattice of nuclei at matter densities
$\rho$ below the
neutron-drip theshold $\rho_{nd} = 4.3\times 10^{11}$ g cm$^{-3}$
or about their stress-induced mobility. 
At densities in the crust above
$\rho_{nd}$, it is likely that neutron star matter is an amorphous
heterogeneous solid, analogous with a disordered many-component
alloy (Jones 1999),
and again, nothing is known about its likely response to periodic
stress.  Thus there is no basis for even a rough estimate of
the contribution to shear-mode $Q$-values from this
mechanism.

Blaes et al assume that coupling with magnetospheric Alfv\'{e}n waves
is the dominant mechanism of damping.  These authors show that, 
for a typical frequency
of $10^{4}$ Hz, shear waves are efficiently reflected from the boundary
with the magnetosphere and, in the absence of other damping
processes, would have a lifetime $\tau_{s}\approx 1$ s at a
reference magnetic flux density of $10^{11}$ G.  The B-dependence
of $\tau_{s}$ is given in analytical expressions valid
asymptotically at
frequencies small and large compared with $10^{4}$ Hz.
These are $\tau_{s}\propto B^{-2}$ at low freqencies and
$\tau_{s}\propto B^{-4/7}$ at high frquencies.
The coupling is
therefore an efficient mechanism for energy transfer from crust to
magnetosphere at the high magnetic fields assumed in magnetars
(Thompson \& Duncan 1995).
Sudden movements of the solid crust, caused by
Maxwell stresses or otherwise, may then lead to magnetospheric
excitation and possibly observable phenomena.  In this context,
the present paper examines some processes, specific to
neutron stars, which are more accessible to calculation than
stress-induced dislocation movement and which might be capable
of damping shear waves enough to invalidate this assumption.
Shear waves are reflected many times from the boundary with the
magnetosphere and possibly from the inner boundary of the crust
with the liquid core of the star where the matter density is $\rho_{c}
\approx 2\times10^{14}$ g cm$^{-3}$.
The number of reflections must be at least
of the order of $c_{s}\tau_{s}/R\approx 10^{2}$, where $R$ is the
neutron star radius and
$c_{s}\approx 10^{8}$ cm s$^{-1}$ is the shear-wave propagation
velocity.  Reflection from each boundary is complicated because it
couples shear waves with longitudinal sound (see, for example,
Landau \& Lifschitz 1970, p.103) and at $\rho_{c}$, may even
depend on the
dimensionality of the solid.

It is known that longitudinal
sound is strongly absorbed in the core of a
neutron star owing to bulk viscosity produced by
the strangeness-changing  four-baryon weak interaction.
The magnitude of the coefficient is very dependent on the
assumptions which must be made, principally about the equation of
state at core densities, but
the absorption is always strong enough, within certain temperature
intervals, to be significant for shear-wave damping.
  
The shear-wave
velocity depends on nuclear parameters according to
$c_{s} \propto ZA^{-2/3}\rho_{N}^{1/6}$, where $Z$ and $A$ are
the nuclear charge and mass number, and
$\rho_{N}$ is the spatially-averaged matter density derived from
the nuclei alone and excluding the superfluid
neutron continuum external to them.
This elementary expression assumes a nuclear effective mass
equal to the free mass and, in conjunction
with the nuclear parameters listed by Negele \& Vautherin (1973),
gives a rapid
increase of $c_{s}$ with depth at $\rho<\rho_{nd}$ but no more
than a very slow increase at $\rho>\rho_{nd}$.  (The latter
variation seems to be in disagreement with Blaes et al who
state that $c_{s}$ is a decreasing function of depth at 
$\rho>\rho_{nd}$.) As a consequence of this variation,
shear waves will be refracted radially
outward in all regions, as are longitudinal sound waves,
but only weakly so at $\rho>\rho_{nd}$. But in a typical
$1.4 M_{\odot}$ neutron star, the
layers of strong refraction at $\rho<\rho_{nd}$ have depth only
of the same order as the wavelength at the relevant frequency of
$10^{4}$ Hz.  Even assuming, with Blaes et al, that
the Maxwell-stress generated
events producing shear waves occur predominantly within this
latter region, it follows from this small depth
and from the multiplicity of reflections
that shear-wave propagation in the neutron drip region at
$\rho>\rho_{nd}$ must be considered as a significant source of damping.
Thus most of the present paper will describe an investigation in
this context of phenomena involving the rapid forced motion of vortices
which, for want of any better name, we shall refer to as
superfluid friction. The relevant results for
longitudinal sound absorption in the liquid core are considered
in Section 3.

\section[]{Description of Superfluid Friction}

An elementary description of superfluid friction is given by
considering a region of solid in which the neutron superfluid
has rotational angular velocity $\Omega_{n}$.  The  
superfluid velocity ${\bf v}_{n}$ is defined (see Sonin 1987)
as the average over the unit cell of the local vortex lattice.
All velocities
are referred to cylindrical polar coordinates rotating with constant
angular velocity $\Omega$, coincident at some fiducial time with the
angular velocity of the charged components of the star.
The difference
$\Omega_{n}-\Omega$ is always small compared with $\Omega$.
To the extent that steady spin-down
of the superfluid occurs, it is maintained by radial outward expansion of
its vortex lattice; the Magnus force ${\bf f}_{M}$ acting on an
individual
vortex is balanced by the force per unit length
${\bf f}$ generated through its motion with velocity $-{\bf u}$,
perpendicular to its local axis and
relative to the solid.  The quantities ${\bf f}$ and ${\bf u}$
have to be treated as averages over macroscopic lengths and times,
an assumption which is satisfactory
provided vortex curvature on these same scales is negligible.
A shear wave of wavevector ${\bf q}$
is represented by a solid matter displacement vector
$\mbox{\boldmath $\xi$}=\mbox{\boldmath$\xi$}_{o}
\exp i({\bf q}\cdot {\bf r}-\omega t)$,
with $\mbox{\boldmath$\xi$}\cdot {\bf q}= 0$.  The velocity
relative to the local
solid is then $-{\bf u}={\bf v}_{L}-{\bf v}_{\perp}$, where the
shear-wave displacement  velocity is
${\bf v}=\mbox{\boldmath$\dot{\xi}$}$ and ${\bf v}_{\perp}$ is its component
perpendicular to the vortex axis. 
The velocity of a point on the vortex axis
is ${\bf v}_{L}$. Energy dissipation
is obviously changed by the displacement velocity, specifically,
by the component ${\bf v}_{\perp}$.
Energy is transferred irreversibly from the shear wave itself
and, for most wave directions and polarizations, from the
rotational kinetic energy of the superfluid measured in the
rotating coordinate system. It is to be expected
that calculation of
shear-wave energy-loss rates depends principally on the
assumptions made about the dissipative force, and so we are
obliged to consider these in as much detail as possible. 

\subsection{The dissipative force}

The assumptions, made here for almost rectilinear vortices, are
that it is
possible to define, averaged over macroscopic lengths
and times, a force ${\bf f}(u)$ per unit length of vortex, and
that fluctuations from it are not so large as to make its use
meaningless.  The
component of this force parallel with ${\bf u}$ is the dissipative
force ${\bf f}_{\parallel}(u)$.  Its properties were considered briefly
in a previous paper (Jones 2002), but it seems appropriate to give
here a more detailed account of what can be deduced about it from
the interaction of a vortex, comoving with superfluid of
velocity ${\bf v}_{n}$, and a single point defect
in the solid structure.
This leads to the excitation of phonons
in the solid and, at finite temperatures, of
vortex-core quasiparticles (Jones 1990a,b),
but Kelvin wave excitation is easily the most important process
(Epstein \& Baym 1992, Jones 1998).
In the case that the interaction potential is attractive,  
the coupling always becomes strong at sufficiently small $v_{n}$ so that
the energy transfer to Kelvin waves must be calculated
non-perturbatively.  In the strong-coupling region, the vortex is
trapped in the potential for a time $t_{m} \propto v_{n}^{-2}$ and
the energy transfer is $E_{a} \propto v_{n}^{-1}$.  We refer to
Jones (1998) for details and for a diagram showing a typical
vortex orbit within the potential. (For weak coupling,
realized only at values of $v_{n}$
larger than those considered here, the energy transfer changes to
$E_{a} \propto v_{n}^{-1/2}$ and the vortex orbit is little perturbed
by the potential.)
Fig. 1 shows schematically
the effect of a sequence of such interactions, with a dilute
distribution of point defects, on the vortex path.
Each interaction produces a lateral displacement and
introduces a trapping time $t_{m}$.
It is found that, averaged over impact parameter, the components
of the computed impulse on the vortex are approximately equal,
\begin{eqnarray}
\left|\int^{\infty}_{-\infty}F_{1}(t)dt\right| \approx
\left|\int^{\infty}_{-\infty}F_{2}(t)dt\right|,
\end{eqnarray}
where the axes are as labelled in the Fig. 1.
The connection between ${\bf F}$, defined for interaction with
an isolated point defect, and the force per unit length
${\bf f}$, which is
an average over macroscopic lengths and times for
motion through a dense system of point defects, is a complex
problem not considered here.
\begin{figure}
\scalebox{0.5}{\includegraphics*[2cm,0cm][19cm,25cm]{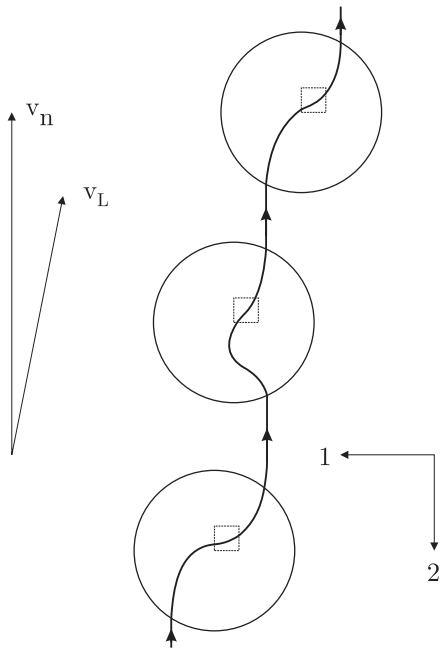}}
\caption{A vortex moves through, and interacts with, a dilute system
of attractive point defects, shown schematically.  In the absence of
interaction, it would comove with the superfluid velocity
${\bf v}_{n}$.  However weak the attractive potential, the orbit
within the potential field, at sufficiently small $v_{n}$, is
always characterized by a mean lateral displacement and by trapping
for finite times in regions close to each origin inside
the small rectangular boxes.
Averaged over macroscopic time and length-scales, a vortex
typically has velocity ${\bf v}_{L}$, as shown.}
\end{figure}
On microscopic length and time-scales,
the vortex has fluctuating motion and, 
as shown in Fig. 1, the macroscopically-averaged vortex velocity
$-{\bf u} = -u\hat{\bf u}$ is smaller in magnitude than the
superfluid velocity ${\bf v_{n}}$ with which the vortex would comove
in the absence of interaction, and has a
component perpendicular to it.  For general values of $v_{n}$ and for
a dense system of point defects, it must be presumed that as a
consequence of equation (1) 
the force ${\bf f}$ has,
in addition to the dissipative component ${\bf f}_{\parallel}$,
a component ${\bf f_{\perp}}$
parallel with $\mbox{\boldmath$\hat{\kappa}$} \times \hat{\bf u}$, where
the circulation is $\mbox{\boldmath$\kappa$} = \kappa
\mbox{\boldmath$\hat {\kappa}$}$ and $\kappa$ is its quantum.
In other superfluid problems,
the existence of this non-dissipative component has been
viewed as controversial.  Thouless, Ao \& Niu (1996) have claimed
that it does not exist for neutral superfluids in a uniform
normal background (see, however,
Hall \& Hook 1998, Sonin 1998, and the replies therein ),
but a system of point defects is not of this class.
However, in the special case of an infinite linear array of point
defects, parallel with $\mbox{\boldmath$\kappa$}$, the Green
function used by Jones (1998) allows one to see that the condition
\begin{eqnarray}
\int^{t}_{-\infty}F_{2}(t^{\prime})dt^{\prime} \rightarrow  0,
\end{eqnarray}
which is necessary for the onset of pinning, can always be
satisfied at sufficiently small $v_{n}$. 

The qualitative form inferred for the dissipative force per
unit length of vortex, $f_{\parallel}(u)$, is shown in Fig. 2.
Finite temperatures and quantum-mechanical tunnelling lead
to vortex creep and therefore
a large and rapidly varying gradient close to $u=0$.  However,
those properties of $f_{\parallel}$ which are important here can be
expressed in the simple function
\begin{eqnarray}
f_{\parallel}(u)= f_{m}\left(1-\rmn{e}^{-u/a}\right)\left
(1-\rmn{e}^{-b/u}\right)
\end{eqnarray}
where $a,b$ and $f_{m}$ are constants.  We assume that $b$ is of
the order of $v_{m}=f_{m}/\rho_{n}\kappa$, where $\rho_{n}$ is the
density of the superfluid continuum outside the nuclei.  It is the
only naturally occurring temperature-independent
scale of velocity in the problem.  The precise value of the parameter
$a$ is not important here except in connection with rotational
noise generation as mentioned in Section 3; it is very many orders
of magnitude smaller than $b$.

The linearized equations for the macroscopic hydrodynamics of an
incompressible superfluid rotating with angular velocity $\Omega$
are:
\begin{eqnarray}
\rho_{n}\mbox{\boldmath$\kappa$}\times({\bf v}_{L}-{\bf v}_{n}) +
f_{\parallel}(u)\hat{\bf u} + f_{\perp}(u)
\mbox{\boldmath$\hat{\kappa}$}\times\hat{\bf u} = 0,
\end{eqnarray}
\begin{eqnarray}
\frac{\partial {\bf v}_{n}}{\partial t} + 2{\bf \Omega}\times
{\bf v}_{L} = -\frac{1}{\rho_{n}}\nabla P,
\end{eqnarray}
where $P$ is the pressure.
Comparison with the equations given by Sonin (1987; equations
4.36 - 4.38) shows that the Tkachenko force, derived from the
vortex lattice structure, and the force dependent on vortex
curvature have been excluded from equation (4).  Both are
quite negligible in the present context.   These equations
are macroscopic in that equation (5) contains the vortex
density $2\Omega/\kappa$, although equation (4), with no
Tkachenko force, is valid for an isolated vortex.  At the
low vortex densities ($10^{3-5}$ cm$^{-2}$) of rotating
neutron stars, $v_{n}$ does not change rapidly with time and,
for the purpose of understanding vortex unpinning and repinning
at a qualitative level, 
can be treated as a constant in equation (4).  Remembering
that ${\bf u} = {\bf v} - {\bf v}_{L}$, and that all the
velocity components with which we are concerned are
perpendicular to $\mbox{\boldmath$\hat{\kappa}$}$, the components of
equation (4) parallel with $\hat{\bf u}$ and
$\mbox{\boldmath$\hat{\kappa}$}\times\hat{\bf u}$ give an equation
for $u$,
\begin{eqnarray}
\left({\bf v}-{\bf v}_{n}\right)^{2} = \left(u-
\frac{f_{\perp}}{\rho_{n}\kappa}\right)^{2} +
\left(\frac{f_{\parallel}}{\rho_{n}\kappa}\right)^{2}.
\end{eqnarray}
The roots of this equation are determined by the $u$-dependence
of the functions $f_{\parallel}$ and
$f_{\perp}$. For consistency with equation (2), the latter
must satisfy the limit $f_{\perp}(u)\rightarrow 0$
as $u\rightarrow 0$.
In general, there can be either one or three roots
in each of the sectors $\mid {\bf v}-{\bf v}_{n}\mid   < v_{m}$
and $\mid {\bf v}-{\bf v}_{n}\mid   > v_{m}$.
A simple example is the instance in which the solid is stationary
in the rotating coordinates so that ${\bf u}=-{\bf v}_{L}$,
with force components 
$f_{\perp}=0$, and $f_{\parallel}$ given by equation (3).
The solutions are shown in Fig. 2
for a set of four values of $v_{n}$.
\begin{figure}
\scalebox{0.5}{\includegraphics*[2cm, 0cm][19cm, 25cm]{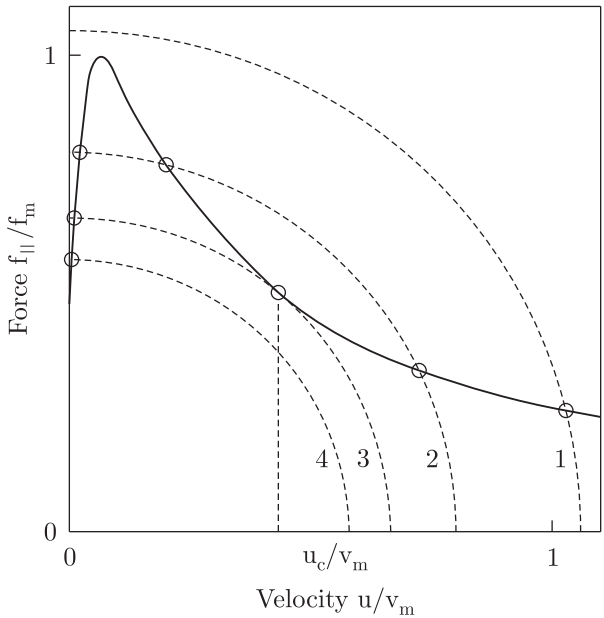}}
\caption{The force per unit length $f_{\parallel}$ (in units of $f_{m}$)
is shown versus $u$ (in units of $v_{m}$).  The function is that of
equation (3) except that the value of the parameter $a$ has been
greatly exaggerated for ease of representation.  For the same
reason, the second parameter is $b = 0.3  v_{m}$, a value intermediate
between those considered in Figs. 3 and 4.  Possible solutions
of equation (6), for the case $v = f_{\perp} = 0$, are given by
the intersection of this function with the circles of radius
$v_{n}/v_{m}$.  For $v_{n} > v_{m}$ (case 1), the vortex
approximately comoves with the superfluid.  With decreasing
$v_{n}$ (cases 2 and 3), the critical value $u_{c}$ is
approached, leading to vortex repinning (case 4)}
\end{figure}
In case (1), the single root
corresponds with an unpinned vortex moving with a velocity
$v_{L}\approx v_{n}$.  If $v_{n}$ were decreased to case (2), there
would be three roots including a pinned solution at very small $u$.
Further reduction to case (3) leads to a critical value
$u=u_{c}$ ($u_{c}= 0.44 v_{m}$ for $b= 0.3 v_{m}$)
and to a transition to the pinned solution.  At the smaller
$v_{n}$ of case (4), only the pinned solution exists.
If equation (6) has only one root, there is, of course, no 
discontinuity and $u$ varies continuously with $v_{n}$.
Obviously, this discussion is based on our definition and use
of macroscopic averaging, but gives some idea of  the unpinning
and repinning processes for an isolated vortex. The very low vortex
densities in neutron matter suggest that these, rather than
multiple-vortex processes involving vortex-vortex interactions,
are the significant consideration.

\subsection{The hydrodynamic equations}

Although the non-dissipative force component $f_{\perp}$ is
expected to be finite for general values of $u$, any attempt
to give it an explicit form would be much less well-based
than equation (3).  Thus the assumption $f_{\perp}=0$ adopted here
is the only practical basis for obtaining numerical
solutions of the hydrodynamic equations.  Resolved into components
parallel with the unit vectors $\hat{\bf r}$ and
$\mbox{\boldmath$\hat{\theta}$}$
in cylindrical polar coordinates, equations (4) and (5) are,
\begin{eqnarray}
f_{\parallel r} - \rho_{n}\kappa\left(v_{L\theta}-v_{n\theta}\right) = 0,
\end{eqnarray}
\begin{eqnarray}
f_{\parallel \theta} + \rho_{n}\kappa\left(v_{Lr} - v_{nr}\right) = 0,
\end{eqnarray}
\begin{eqnarray}
\dot{v}_{nr} = 2\Omega v_{L\theta} - \frac{1}{\rho_{n}}\nabla_{r}P,
\end{eqnarray}
\begin{eqnarray}
\dot{v}_{n\theta} = -2\Omega v_{Lr} - \frac{1}{\rho_{n}}
\nabla_{\theta}P.
\end{eqnarray}
With regard to boundary conditions, the only practical way of
obtaining numerical solutions is to assume an infinite system.
Thus the boundary condition that ${\bf v}_{n}$ should
have no perpendicular component on the 
spherical surface of the superfluid at $\rho_{nd}$ is
neglected.  Solutions are therefore valid only for
wavenumbers $q \stackrel{>}{\sim}10^{-4}$ cm$^{-1}$ large enough
to be consistent with the infinite system assumption.
Equations (7)-(10) also neglect fluctuations in pressure and
gravitational potential, a reasonable approximation
at the wavenumbers considered here.
Any pressure fluctuation caused by
the local shear
displacement is negligibly small and it is therefore
appropriate to adopt the initial conditions $\nabla_{r}P =
\nabla_{\theta}P = 0$.  It will be convenient to eliminate
${\bf v}_{L}$ from equations (7)-(10) in favour of ${\bf u}$.
We also neglect any small differences between the local
$\mbox{\boldmath$\hat{\kappa}$}$ and the $\hat{z}$ axis.
The solid velocity ${\bf v} = {\bf v}_{o}\sin{\omega t}$
is resolved into components parallel
and perpendicular to the $\hat{z}$ axis, only the latter,
${\bf v}_{\perp}$ being
present in equations (7)-(10).
The velocity variables are then
$u_{r}, u_{\theta}, v_{nr}$ and $v_{n\theta}$, with initial
conditions of $v_{nr}(0)=0$ and a specified value of
$v_{n\theta}(0)$.  The initial values, $u_{r,\theta}(0)$, obtained
from equations (7) and (8) are for a pinned-vortex solution
which is always assumed as the initial condition.
Algebraic solution of equations (7) - (10) is possible
through linearization, which is valid for shear waves of such small
amplitude that they cannot cause vortex unpinning. 
It gives a dissipation rate per unit length of vortex
of the order of
\[\left(\rho_{n}\kappa v_{o \perp}\right)^{2}
\left(\frac{\partial f_{\parallel}}{\partial u}\right)^{-1}.\]
For any physically reasonable
values of the parameters concerned, this is
many orders of magnitude too small to be of interest
in the present context.

The rate of energy dissipation
per unit length of vortex is ${\bf f}\cdot{\bf u}$, which satisfies
\[{\bf f}\cdot{\bf u} = {\bf f}\cdot{\bf v}-{\bf f}\cdot{\bf v}_{L}.\]
Because ${\bf f}+{\bf f}_{M} = 0$ and
${\bf f}_{M}\cdot({\bf v}_{L}-{\bf v}_{n}) = 0$, this can be rewritten
as
\begin{eqnarray}
{\bf f}\cdot{\bf u} = {\bf f}\cdot{\bf v} - {\bf f}\cdot{\bf v}_{n}.
\end{eqnarray}
This merely states that the dissipation rate is the sum of the
energy loss rates for the shear mode (${\bf f}\cdot{\bf v}$) and for
the superfluid.  For the chosen initial conditions, equations
(7)-(10) have been solved numerically, with attention to the
instabilities discussed in Section 2.1, for several complete
cycles of the shear mode in order to obtain values of its time-averaged
energy loss rate.  These rates have been calculated for the interval
$10^{4} < v_{o \perp} < 10^{6}$ cm s$^{-1}$, where ${\bf v}_{o \perp}$
is the component of the displacement velocity ${\bf v}$
perpendicular to $\mbox{\boldmath$\kappa$}$, and for a distribution
of values of its $r,\theta$ components. This allows
us to obtain, for a given value of $v_{o}$ and for each of the shear
mode polarization states, the energy loss rate
averaged over an isotropic distribution of its wave vector
${\bf q}$. (The multiple reflection of shear waves makes this average
the most appropriate way of presenting the data.) The energy loss-rates
per unit length of vortex, divided by $\rho_{n}\kappa$ and
averaged over both polarization states, are defined as
$D_{av}$ and shown in Fig. 3 as
functions of $v_{o}$ for pairs of values of $f_{m}/ \rho_{n}\kappa$
and of $b$. They are quite insensitive to $\omega$, $\Omega$ and $a$.
Therefore, we have assumed $\omega = 10^{4}$ rad s$^{-1}$,
$\Omega = 10$ rad s$^{-1}$ and $a = 10^{-4}$ cm s$^{-1}$ throughout.
Except for velocities close to the unpinning threshold,
loss-rates do not depend strongly on the initial pinned-vortex velocity.
This is demonstrated by the broken curve in Fig. 3 which is for
$u_{r}(0) = 0$, an initial state of an exactly corotating superfluid,
$v_{nr} = v_{n\theta} = 0$.
Away from the threshold, it is almost indistinguishable
from the curve for $u_{r}(0) = a$ and so we have assumed this latter
initial condition for all other curves shown.
In general, energy loss-rates
become large at values of $v_{o}$ sufficient to produce vortex unpinning
and increase to asymptotic values
\begin{figure}
\scalebox{0.5}{\includegraphics*[2cm, 0cm][19cm, 27cm]{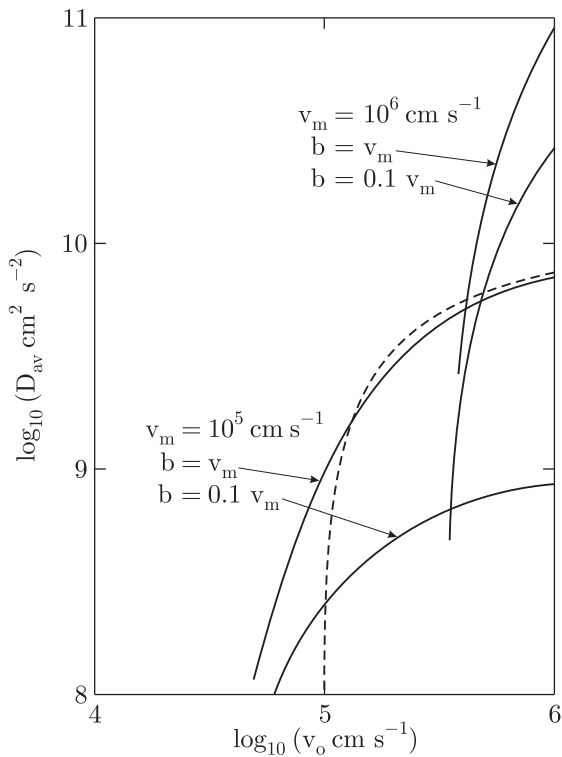}}
\caption{The shear-mode energy loss-rate per unit length of vortex,
averaged over polarization and for an isotropic distribution of
wavevectors, is shown as a function of velocity amplitude $v_{o}$
for given $v_{m}$ and $b$.  The broken curve, for $b = v_{m} =
 10^{5}$ cm s$^{-1}$, is for the initial condition
$v_{nr} = v_{n \theta} = 0$.}
\end{figure}
\begin{figure}
\scalebox{0.5}{\includegraphics*[2cm, 0cm][19cm, 27cm]{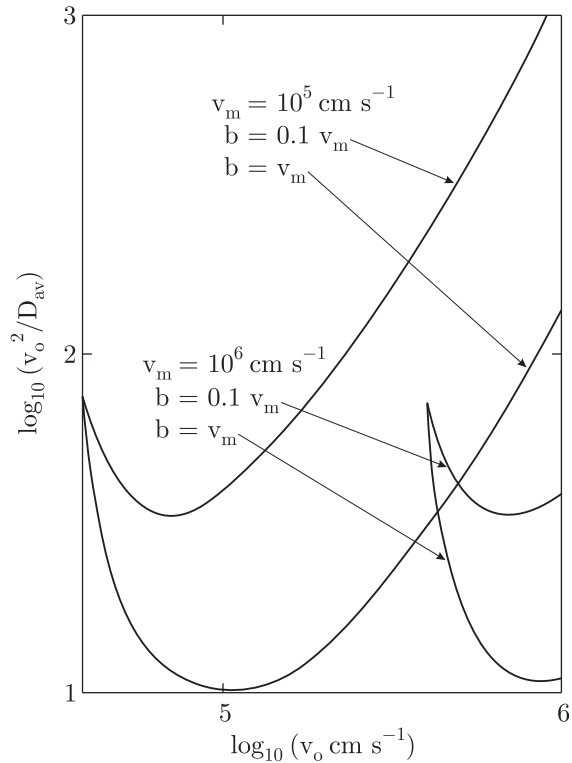}}
\caption{The vertical axis gives the
right-hand side of equation (12), which is
proportional to $\Omega\tau$
averaged over polarization and for an isotropic
distribution of wavevectors.  The
parameter sets are those of Fig. 3.  Damping is strongest at values of
$v_{o}$ in the vicinity of $v_{m}$.}
\end{figure}
$\rho_{n}\kappa D_{av} \approx f_{m}b$.
The mode decay time $\tau$ at a given amplitude $v_{o}$ can be found from
its energy density $\frac{1}{2}\rho_{N}v^{2}_{o}$ and the loss-rate per
unit volume, equal to the product of $\rho_{n}\kappa D_{av}$ and the
vortex number density $2\Omega / \kappa$.  The function
\begin{eqnarray}
\frac{v^{2}_{o}}{D_{av}} = 4\Omega\tau\frac{\rho_{n}}{\rho_{N}}
\end{eqnarray}
is shown in Fig.4. Assuming the lattice parameters calculated by
Negele \& Vautherin (1973),
the density ratio $\rho_{n}/\rho_{N}$ appearing
in our, perhaps naive, model of the shear wave varies from zero at
the threshold $\rho_{nd}$ to
approximately 8 in the interval $10^{13} < \rho < 10^{14}$ g cm$^{-3}$
which includes most of the neutron-drip volume. The product
$\Omega \tau$ is less than unity for a substantial interval
of shear-wave amplitude.

\section[]{Discussion and conclusions}

The motivation for expressing the force parameter $f_{m}$ as in
Figs. 3 and 4 is that $v_{m} = f_{m}/\rho_{n}\kappa$ is an upper
limit for the superfluid velocity $v_{n\theta}$ relative to
coordinates rotating with angular velocity $\Omega$.  Thus values
of the order of $10^{4}-10^{6}$ cm s$^{-1}$ must exist in some
intervals of matter density if angular momentum transfer from crust
superfluid is the origin of, or a substantial contributor
to, pulsar glitches.  Much smaller values
may exist in other regions, and the limit $v_{m}\rightarrow 0$
must be approached  near $\rho_{nd}$ and $\rho_{c}$. In Section 2.1,
it was argued that $b$ should be of the order of $v_{m}$, but
in Figs. 3 and 4, results are also given for values an order
of magnitude smaller. From Fig. 3, we can see that the asymptotic
value $D_{av}\approx v_{m}b$ gives a very approximate estimate,
$2\Omega\rho_{n}v_{m}b\theta(v_{o}-v_{m})$, for the energy-loss rate
per unit volume, where $\theta$ is here the unit step function.
The order of magnitude of the energy decay time is then
\begin{eqnarray}
\tau = \frac{\rho_{N}v_{o}^{2}}{4\rho_{n}\Omega v_{m}^{2}
\theta(v_{o}-v_{m})},
\end{eqnarray}
for $b=v_{m}$.  It is a function only of $v_{o}/v_{m}$, so that
even a small-amplitude shear wave is strongly damped in regions
of weak vortex-pinning. It is also temperature-independent,
below the superfluid critical temperature,
because it depends principally on $f_{\parallel}$ at values of $u$
many orders of magnitude greater than $a$.  The conventionally
defined shear viscosity $\eta$ for the solid (see Landau \&
Lifshitz 1970, p. 155) is related with the decay time by
$\eta = \mu/\tau\omega^{2}$, where $\mu$ is the shear modulus.  

This process has some similarity with stress-induced motion of
dislocations, considered briefly by Blaes et al.  Both involve
dissipative motion of a topological defect, either
in the superfluid order parameter or in the lattice, but it
is not possible to obtain even an order of magnitude
estimate of the importance of the latter process.  The force per
unit length on a line dislocation is of the order of
$\tilde{b}\mu v_{o \perp}/c_{s}$, where
$\tilde{b}$ is the Burgers vector (see Landau \& Lifshitz 1970,
p. 133); it is typically some orders
of magnitude smaller than $\rho_{n}\kappa D_{av}$. The dislocation
density is not known, but could be many orders of magnitude greater
than the vortex  density, $2\Omega/\kappa \approx 10^{3-5}$ cm$^{-2}$.
But the time-averaged velocity of an individual dislocation
under stress is
probably very many orders of magnitude smaller than $v_{m}$.

The coupling with longitudinal sound referred to in Section 1
introduces additional damping.  The most significant source is
the bulk viscosity produced by the strangeness-changing
four-baryon weak interaction.  The amplitude decay time is
\begin{eqnarray}
\tau_{nl} = \frac{2\rho c^{2}_{\ell}(0)}{\zeta\omega^{2}},
\end{eqnarray}
at matter density $\rho$ and angular frequency $\omega$, where
$c_{\ell}(0)$ is the longitudinal sound velocity in the
zero-frequency limit.  The bulk viscosity coefficient $\zeta$ is
a function both of temperature and angular frequency. It also
depends on the equation of state
at core densities.  For recent evaluations, based on differing
assumptions, we refer to Jones (2001), Haensel et al (2002) and
Lindblom \& Owen (2002).  For $\omega=10^{4}$ rad s$^{-1}$,
a temperature of $10^{9}$K and
$\zeta=3\times 10^{28}$ g cm$^{-1}$s$^{-1}$
(see Jones 2001; equation 45), the decay time is
$\tau_{nl} = 3\times10^{-2}$ s, which would
increase by several orders of magnitude if the temperature were to
decrease to $10^{8}$K.  The evaluation of Lindblom \& Owen gives a
larger $\zeta$ and thus an even shorter decay time.

Fig. 3 of Blaes et al shows the computed Alfv\'{e}n wave luminosity
as a function of time following generation of shear waves by a
starquake, assuming a magnetic flux density equal to their
reference value of $10^{11}$ G.
  Energy begins to enter the magnetosphere about $10^{-3}$s
after the event and the luminosity persists for times just
exceeding $6\times 10^{-1}$s.  Anomalous X-ray pulsars (AXP) and
soft gamma repeaters (SGR) are associated with long rotation periods,
$\Omega \approx 1$ rad s$^{-1}$ and, in the magnetar model,
with surface magnetic fields exceeding $10^{14}$ G. These very
large fields indicate,
assuming the $B$-dependence of $\tau_{s}$ given by Blaes et al
and referred to in Section. 1,
that superfluid
friction damping times are probably rather longer than
the time $\tau_{s}$ for energy
transfer by coupling of the shear modes with magnetospheric
Alfv\'{e}n waves. In magnetars, superfluid friction is unlikely
to limit the
efficiency of this process.  A similar
conclusion, though depending on temperature, holds for the
bulk viscosity damping.

Duncan (1998) has
examined the excitation and detectability of global seismic modes,
particularly low-order toroidal modes, by starquake events.  He
assumes that these modes are damped through their interaction with
fault lines in the solid and the consequent generation of high-frequency
shear waves.  But the superfluid friction calculated in the present
paper damps such modes, based on shear deformations, directly. The
damping times found here are much shorter than those considered by
Duncan and would rule out the existence of other than small
mode-amplitudes.

The presence of $\Omega$ in equations (12) and (13) represents no
more than the mean vortex density of the superfluid.  Therefore,
superfluid friction would produce extremely strong damping in
more rapidly rotating neutron stars, such as the typical radio
pulsar with $\Omega \approx 10^{1-2}$ rad s$^{-1}$.  This may
also be of relevance to the
AXP and SGR sources,
of which all examples so far observed have long periods. Energy
transfer through coupling between shear waves and magnetospheric
Alfv\'{e}n waves, as in the magnetar model of Thompson
\& Duncan (1995), is unlikely to be efficient at short periods
owing to the very strong damping.

Superfluid friction is also likely to be relevant to other current
problems such as the possibility of vortex unpinning in precessing
neutron stars (Link \& Cutler 2002), and the damping of r-modes in
rapidly rotating neutron stars
(Lindblom, Owen \& Morsink 1998; Andersson, Kokkotas \& Schutz 1999).
Many papers on the latter problem have assumed that the solid crust
does not participate in the displacement amplitude of the mode, is
stationary in rotating coordinates, and therefore acts only as a
boundary
condition on a thin viscous or turbulent dissipative boundary layer.
But more recent work (Levin \& Ushomirsky 2001; Yoshida \& Lee 2001)
has recognized that the mode amplitude must penetrate the crust
significantly at high values of $\Omega$. (The radial component
of the r-mode displacement, though small, must penetrate the crust.
Also, an element of solid crust
moving with velocity ${\bf v}$ in the rotating coordinates
experiences a Coriolis force which, for $\Omega \approx 10^{2}$ rad
s$^{-1}$, is of the same order of magnitude as the shear stress
integrated over its surface.) A mode with amplitude large enough for
its velocity of displacement to be of the order of $v_{m}$ will
be subject to strong dissipation in the crust, the energy
loss-rate there being of the order of $2\Omega\rho_{n}v_{m}b
\theta(v-v_{m})$ per unit volume, as for shear
waves.  Unlike kinetic theory or weak-interaction mechanisms
of viscous dissipation (Jones 2001; Haensel et al 2002;
Lindblom \& Owen 2002), this is
not cut off by exponential factors involving superfluid energy gaps
but is approximately temperature-independent below the critical
temperatures.
However,the detailed calculation required to obtain the
r-mode lifetime is beyond the scope of the present paper.

The second term on the right-hand side of equation (11) is the
energy-loss rate for the neutron superfluid.  For all the cases
in Fig. 3, except
for that with the initial condition $u_{r}(0) = 0$, the superfluid
loses kinetic energy in the rotating frame of reference and is
gradually brought into corotation with the charged components
of the star.  The initial rate is of the same order as for the
shear mode.

Superfluid spin-down in the presence of shear waves is governed by
the form of $f_{\parallel}(u)$ at $u < u_{m}$, the position
$u_{m}$ of its maximum being 
$u_{m} \approx \sqrt{ab}$ for the specific form given by equation (3).
In the absence of shear waves, the solid matter velocity is $v = 0$
and steady-state superfluid spin-down is possible, in principle,
if the required vortex-creep velocities,
$v_{L}= -r\dot{\Omega}/2\Omega$ at radius $r$, are smaller than
$u_{m}$.  The actual form of $f_{\parallel}$ in this region is
unknown, because it is dependent on thermal activation and
quantum-mechanical
tunnelling, but it would not be unreasonable to assume that it allows
this condition to be satisfied given the very small values,
$v_{L} < 10^{-5}$ cm s$^{-1}$ required.  Under this assumption, some
perturbing movement of the solid is needed to induce vortex unpinning.
Its magnitude depends on the actual form of $f_{\parallel}(u)$
near its maximum at $u_{m}$, but it is likely that quite small values
would be sufficient to satisfy the condition
$\mid {\bf v} - {\bf v}_{n} \mid > v_{m}$ in one or more,
possibly small, regions of superfluid.  Following our
discussion at the end of Section 2.1, we can see that
vortices in these regions would remain unpinned with rapid
superfluid spin-down, even if
the initial disturbance were to decay to negligible amplitudes,
until the critical value $u = u_{c}$ is reached. Thus a
significant fraction,
$1-\sqrt{(u_{c}/v_{m})^{2}+(f_{\parallel}(u_{c})/f_{m})^{2}}$,
of the available superfluid angular momentum in these regions is
transferred to the charged components of the star in discontinuous
glitch or noise events.

It is worth comparing the crust movements considered here with
those assumed in the glitch model of Ruderman, Zhu \& Chen (1998).
The poloidal velocities of the latter model are not necessarily
large enough to generate a significant shear-wave energy density;
it is possible that they are no more than episodes of plastic
flow occurring in times smaller than the observed upper limit
($\approx 10^{2}$s) for pulsar spin-up, thus with velocities
many orders of magnitude smaller than the shear-wave velocity
$c_{s}$.  In this case, superfluid
is spun up or down by the inward or outward radial displacement
of the solid with pinned vortices (see Jones 2002).  Vortices are
not necessarily unpinned although this could occur in any region
where the superfluid velocity $v_{n}$ reaches $v_{m}$.  But crust
movement with poloidal velocities of the order of $c_{s}$ is not
excluded, in which case glitches would be initiated by the
($v \neq 0$) velocity perturbations considered here.  It is
possible that both kinds
of movement contribute to the observed glitch population.
    
 This would not be inconsistent with the value of the
glitch activity parameter defined by Lyne et al (2000), which is
based on the observation that for all except
very young or old pulsars, the summed glitch amplitudes
$\delta\Omega_{gi}$ divided
by the total observation time is $T^{-1}\sum_{i}\delta\Omega_{gi}
\approx -1.7\times 10^{-2}\dot{\Omega}$.   For this group of
pulsars, angular momentum of the same order as that of the
total crust superfluid is transferred by glitches rather
than by the steady-state superfluid spin-down described
earlier in this paragraph.  Thus all glitches
in radio pulsars may be initiated by
crust movements, with poloidal velocities of the order of
$c_{s}$ or much smaller. Such movements occur less
frequently, if at all, in
very young or old neutron stars for which steady-state 
superfluid spin-down is more usual.
These modes of glitch formation may well extend to very small events,
so giving a model for the observed timing noise in the residual
phase and its time derivatives.

\bsp

\label{lastpage}

\end{document}